 \documentstyle[12pt,fleqn]{article}
\begin{document}
\title{ Generation of a quantum integrable class of discrete-time
 or relativistic periodic  Toda chains}
\author{
Anjan Kundu  \\
Physikalisches Instituet
der Universitaet Bonn\\
Nussallee 12,
53115 Bonn,
Germany}
\date{}
\maketitle
\begin{abstract}
%------------------------------------------------------------
A new integrable  class of quantum models representing
a family of  different discrete-time  or relativistic generalisations of the
periodic Toda chain (TC),
 including that of a recently proposed classical close to TC  model
[7] is presented.
All  such models are shown to be obtainable from a single
ancestor model at different realisations of the underlying quantised algebra.
As a consequence the $(2\times 2)$ Lax operators and the associated quantum
$R$-matrices for these models are easily derived
ensuring their  quantum integrability.
 It is shown
that  the functional Bethe ansatz  developed for the   quntum TC
is trivially generalised to achieve  seperation
of variables also for the present models.

\end{abstract}
\vskip 2 cm
\hrule
{\it Address after June'94}:
 Saha Institute of Nuclear Physics,AF/1 Bidhan Nagar,Calcutta 700 064,India.

\newpage
%%%%%%%%%%%%%%%%%%%%%%%%%%%%%%%%%%%%%%%%%%%%%%%%%%%%%%%%%%%%%%%%%%
%\section{}
\setcounter{equation}{0}
1.~ Toda chain (TC), which is considered to
be one of the most fascinating   integrable lattice systems,
 attracted continuous attention over the years to investigate both its
 classical [1] and
quantum [2,3] aspects.
In recent past a descrete-time  TC (DTTC) was proposed by Suris
at the classical level [4], which was also shown to be equivalent to the
relativistic TC of Ruijsenaars [5]. Following that
we  had obtained [6]
two different quantum generalisations of such DTTC along with the
associated quantum $R$-matrices. Recently another close to TC  model,
 which is classically integrable, has been
put forward [7].
We present here a descrete-time as well as  quantum generalisation
of this  model. However the basic aim of this letter is to
find a new quantum integrable  class  of different  discrete-time
 generalisations
of the  periodic   TC, which are also
canonically related to the  relativistic TC of Ruijsenaars. We show that
 all such models
can be generated  from a single ancestor model
at different realisations of the underlying quantised algebra,
which is  defined
by the
 extended trigonometric Sklyanin algebra  [6]. At the same time we also
 obtain automatically  the
associated quantum $R$-matrix, while the  integrability
condition given by the quantum
Yang-Baxter equation
(QYBE)
\begin {equation}
R(\lambda , \mu)~ L_1(\lambda)~ L_2(\mu )
{}~ = ~  L_2(\mu )~ L_1(\lambda)~ R(\lambda , \mu).
\end {equation}
%1
 with $~ L_1 \equiv
L
\otimes {\bf 1} ,~ L_2 \equiv {\bf 1} \otimes L ~ $
is satisfied by construction. Moreover, we observe
 that the functional Bethe ansatz
 developed
by Sklyanin [3] for the seperation of variables
 in  quantum TC
can  be applied in a parallel way also to its descrete-time generalisations.

Since our aim here is to generate models, we  start not from any particular
model, but directly from  the  quantum $R$-matrix  taking  it
 in the form
\begin {equation}
R_{\theta}(\lambda_{12}) = \left( \begin{array}{c}
 \sin (  \lambda_{12}  +  \alpha ) \ \ \  \qquad \ \qquad \ \qquad \\
    \quad \  \quad e^{-i\theta} \sin  \lambda_{12} \   \quad  \quad
 \sin  \alpha \ \quad  \\
     \quad \  \sin   \alpha \   \qquad  \quad e^{i\theta}
\sin  \lambda_{12}   \ \quad \\   \qquad
        \quad \ \qquad \ \qquad \
 \sin (  \lambda_{12}  +  \alpha )  \end{array}   \right)
\end {equation}
%2
with $ \lambda_{12} = \lambda- \mu$.
Note that (2) is obtained  from the
 standard trigonometric $R_0$-matrix [8] by a simple 'gauge transformation'
 [9]:
$
R_{\theta}= F(\theta)
R_0 F(\theta)  $  with $  F(\theta)=
 e^{i\theta(\sigma_3 \otimes 1-1\otimes \sigma_3)}$, which
takes a solution of the  YBE to another solution of it introducing
an extra  parameter $\theta$.
Now following the
Yang-Baxterisation  scheme for the construction of   Lax operators
[10] one obtains its generalised form
\begin {equation}
L(\lambda) = \left( \begin{array}{c}
\xi {\tau_1^-} + \frac {1}{\xi}{\tau_1^+} \qquad    {\tau_{21}} \\
   {\tau_{12}} \qquad \xi {\tau_2^-} +\frac {1}{\xi} {\tau_2^+}
          \end{array}   \right)
\end {equation}
%3
with spectral parameter  $\xi=e^{i
\lambda}$
 and  yet undefined abstract operators $\vec \tau $.
 The integrability
 condition  on (3), that is the validity of
QYBE (1)
with    $R$-matrix (2), dictates the algebra for $\vec \tau $ as
\begin {eqnarray}
 e^{i\theta} \tau_{12} \tau_{21} - e^{-i\theta} \tau_{21} \tau_{12}
  ~=~
 -2i \sin \alpha \left (  \tau_1^+ \tau_2^-  -  \tau_1^- \tau_2^+ \right )
 ~, \nonumber \\
\tau_i^{\pm }\tau_{ij} ~=~e^{i(\pm \alpha+\theta)}
 \tau_{ij} \tau_i^{\pm }~,~\ \
\tau_i^{\pm }\tau_{ji} ~=~e^{i(\mp \alpha-\theta)} \tau_{ji} \tau_i^{\pm }~,~
\end {eqnarray}
%4
with $i,j = 1,2 $.
We call this algebra extended trigonometric Sklyanin algebra, which may be
reduced to the well known quantum algebra $U_q(su(2))$ at some
particular reduction, though in general, it allows more freedom
necessary for the construction of   integrable models [6]. Thus (3)
represents the Lax operator of a generalised quantum integrable model,
entries of which are governed by  algebra  (4).
\vskip 1 cm
2.~ Now, in order to  generate  a family of DTTC
 related to  $R$-matrix (2),
we consider   different realisations of  algebra (4)
in canonical
 operators $u,p$  with
$[u,p]= i\hbar$.
  First imposing the reduction
$\tau_2^{\pm}=0 $ in (4)  we simplify it to
\begin {eqnarray}
  \tau_{12} \tau_{21}& =& e^{-2i\theta} \tau_{21} \tau_{12}
{}~, \ \nonumber \\
\tau_1^{\pm }\tau_{12} ~&=&~e^{i(\pm \alpha+\theta)}
 \tau_{12} \tau_1^{\pm }~,~\ \
\tau_1^{\pm }\tau_{21} ~=~e^{i(\mp \alpha-\theta)} \tau_{21} \tau_1^{\pm }~
\end {eqnarray}
%5
Observing that only Wyel type relations are involved in the above algebra
we   easily find a
  realisation \begin {eqnarray}
\tau_1^{+ }=e^{(\eta+\epsilon) p} , \ \tau_1^{- }=-e^{-(\eta-\epsilon) p },
\ \
  \tau_{21}=\eta e^{-c \epsilon p+q}, \  \tau_{12} = -\eta e^{(2+c)
 \epsilon p-q
}\end {eqnarray}
%6
 consistent with  (5), where
$\alpha=\eta \hbar, \theta= \epsilon \hbar$.
Since we are dealing here with operators, a certain ordering should
be maintained in (6) and in all other relevant expressions in what
follows and we also consider only the  periodic boundary condisions
for all models.
 Inserting (6) with reduction
$\tau_2^{\pm}=0 $ in  (3) one gets  the
explicit form
\begin {equation}
L_{(\epsilon,c)}(\lambda) = \left( \begin{array}{c}
  \frac {1}{\xi}e^{(\eta+\epsilon) p}
-\xi e^{-(\eta-\epsilon) p}\qquad   \eta e^{-c \epsilon p +q}
 \\-\eta e^{(2+c) \epsilon p-q
}
 \qquad \quad 0
          \end{array}   \right).
\end {equation}
%7
Therefore  Lax operator (7) associated with quantum $R$-matrix (2)
represents an integrable model, which   depends  on
additional parameters $\epsilon$
and $c$. Defining $I^+ \equiv -C_{n-2}(C_n)^{-1}$ and
 $I^+ \equiv -C_{-(n-2)}(C_{-n})^{-1}$, where  $C_{\pm n}$ are conserved
quantities  obtained as the   expansion
coefficients   of the related transfer
matrix
\begin {equation}
t(\xi)=tr T(\xi)=tr (\prod L_n) =tr
\left( \begin{array}{c} A(\xi)
\qquad   B(\xi)
\\
C(\xi) \qquad D(\xi)
          \end{array}   \right)=
 \sum_n C_n\xi^n+C_{-n}\xi^{-n}
\end {equation}
%8
 one arrives at  the Hamiltonian $H= \frac {1}{2}(I^++I^-)
$ and the  momentum
$P= \frac {1}{2}(I^+-I^-)$ of the system
as
\begin {eqnarray}
H=\sum_i\left(\cosh 2\eta p_i
+\eta^2 \cosh \eta (p_i+p_{i+1})e^{\epsilon ( 1+c)(p_{i+1}-p_i)+(q_i-q_{i+1})}
\right),\\
P=\sum_i\left(\sinh 2\eta p_i
+\eta^2 \sinh \eta (p_i+p_{i+1})e^{-\epsilon ( 1+c)(p_i-p_{i+1})+(q_i-q_{i+1})}
\right).
\end {eqnarray}
%9,10
To clarify the meaning of the parameters entering into the   system
 notice that, the Lax operator (7)
and the Hamiltonian (9) depend on the deformation parameters
 $\eta, \epsilon$ along with $c$ coming from the realisation
 (6), but not on  the  parameter $\hbar$.
 On the other hand  $R$-matrix (2) and the quantised algebra (4)
are free from $c$ and depend on
$\alpha=\eta \hbar, \theta= \epsilon \hbar$
, i.e. they depend
not only on  $\eta, \epsilon$, but also on the quantum parameter
$\hbar$ coming from the commutators.
 When  the deformation parameter
  $q=e^{ \eta}\rightarrow 1$, i.e.
$\eta\rightarrow 0$ with finite $\hbar$,
one has to   scale also the
spectral parameters like $\lambda =\eta u$, which
would  reduce the   $R$-matrix  to its  rational form
and the $L$ operator
 to the time-continuous  quantum TC model, though
 more generalised  to include  parameters $\epsilon$ and $c$.
However we notice that for general values of $\theta$  (2)
does not give the classical $r$ as in (12). Therefore such integrable
models interestingly seem to be living  only  at the quantum level.

In the case when $c=-1$ or $\epsilon \rightarrow 0$
one recovers at
$\eta\rightarrow 0$ the
standard Toda chain and  therefore we concentrate now on them.
 Consider first the choice
 $\epsilon = c_0 \eta$ yielding
 from (6):
\begin {eqnarray}
\tau_1^{+ }=e^{\eta(1+c_0) p} , \ \tau_1^{- }=-e^{-\eta(1-c_0) p },
\ \
  \tau_{21}=\eta e^{-\eta c_0 c  p+q}, \  \tau_{12} = -\eta e^{\eta(2+c)c_0
  p-q
}\end {eqnarray}
%11
with $
\tau_2^{\pm }=0$
, which readily gives the Lax operator from (3) associated with $R_\theta$
matrix (2) with $\theta=c_0 \alpha$.
The  classical limit exists  for this model, since the $R$-matrix
at
$\hbar \rightarrow 0$:
\begin {equation}
R(c_0 \alpha, \alpha,\lambda_{12})= I+\hbar
r(\epsilon,\eta, \lambda_{12}) + O (\hbar)\end {equation}
%12
 yields the classical $r$-matrix and algebra (4) with $\theta=c_0 \alpha$
reduces to a Poisson
bracket algebra.
 The relevant conserved charges of the model
may be given by

\begin {eqnarray}
I^{\pm}=\sum_i\left(e^{\pm 2\eta p_i}
+\eta^2 e^{\pm  \eta (p_i+p_{
i+1})+\eta c_0 ( 1+c)(p_{i+1}-p_i)+(q_i-q_{i+1})}
\right).
\end {eqnarray}
%13
Note that at the continous-time  or equivalently at the nonrelativistic
limit
: $\eta\rightarrow 0$
, the corresponding Hamiltonian  $H= \frac {1}{2}(I^++I^-)$
, Lax operator and the   quantum
$R$-matrix yield exactly those belonging to the standard quantum TC.
Thus this model for different values of parameters $c_0$ and
$c$, generates  a  family  of  discrete-time or relativistic  quantum
 integrable Toda chains. Interestingly, since the
$R$-matrix is independent of  parameter $c$, all the models with different
values of
$c$ will share the same quantum $R$-matrix.
Some particular models of  the above integrable class (11,13)
 deserve  special attention
and we will look more closely at them.

For example, at  $c_0=1$ and $c=0$,
(11) reduces to
\begin {eqnarray}
\tau_1^{+ }=e^{2\eta p} , \ \tau_1^{- }=- 1
\ \
  \tau_{21}=\eta e^{q}, \  \tau_{12} = -\eta e^{2\eta
  p-q
}\end {eqnarray}
%14
and the conserved quantities (13)  take  simpler
form as
\begin {eqnarray}
I^{+}=\sum_i\left(e^{ 2\eta p_i}\left(
1+\eta^2 e^{
(q_{i-1}-q_i)}
\right)\right),\quad
I^{-}=\sum_i\left(e^{ -2\eta p_i}\left(
1+\eta^2 e^{
(q_i-q_{i+1})}
\right)\right).
\end {eqnarray}
%15
We immediately recogise this case to be the quantum generalisation
of the DTTC due to Suris [4]. The
associated quantum
$R_{\alpha}$-matrix is obtained easily by putting $\theta =\alpha$ in (2),
 which through (12) recovers exactly the asymmetric classical $r$-matrix
of [4].

A special  case having  the most simplification is obtained when
$c=c_0=0$. As it is seen  from (11) and (13)
 the Lax operator of this model is given by
\begin {equation}
L_{0}(\lambda) = \left( \begin{array}{c}
  \frac {1}{\xi}e^{\eta p}
-\xi e^{-\eta p}\qquad   \eta e^{q}
 \\-\eta e^{-q
}
 \qquad \ 0
          \end{array}   \right),
\end {equation}
%16
yielding the Hamiltonian and momentum in  more symmetric form
\begin {eqnarray}
H=\sum_i\left(\cosh 2\eta p_i
+\eta^2 \cosh \eta (p_i+p_{i+1})e^{q_i-q_{i+1}}
\right),\\
P=\sum_i\left(\sinh 2\eta p_i
+\eta^2 \sinh \eta (p_i+p_{i+1})e^{q_i-q_{i+1}}
\right).
\end {eqnarray}
%17,18
The corresponding $R$-matrix is clearly the $\theta=0$ expression of (2), i.e.
the standard one [8]

\begin {equation}
R_{0}(\lambda_{12}) = \left( \begin{array}{c}
 \sin (  \lambda_{12}  +  \alpha ) \  \qquad \qquad \ \qquad \ \quad \\
    \quad \  \sin   \lambda_{12} \   \quad  \quad \sin \alpha \ \quad  \\
     \quad \  \sin  \alpha \
  \quad  \quad \sin  \lambda_{12}   \ \quad \\
         \qquad \qquad \ \qquad \ \quad \
 \sin (  \lambda_{12}  +  \alpha )  \end{array}   \right).
\end {equation}
%19
Note that  (14) and (16) are the cases  presented in [6].

Another situation of interest appears when $c_0=1$ but $c \not = 0$, which
generalises the quantum Suris model (15)  to give
\begin {eqnarray}
I^{+}=\sum_i\left(e^{ 2\eta p_i}\left(
1+\eta^2 e^{\eta c
(p_i-p_{i-1})+
(q_{i-1}-q_i)}
\right)\right),\nonumber \\
I^{-}=\sum_i\left(e^{ -2\eta p_i}\left(
1+\eta^2 e^{\eta c
(p_{i+1}-p_i)+
(q_i-q_{i+1})}
\right)\right),
\end {eqnarray}
 %20
though sharing the same quantum $R$-matrix with it. The Lax operator
is obtained from (7) by putting $\epsilon=\eta$.

Coming now to  the $c=-1$ case for arbitrary values of $\epsilon$,
it is  seen
clearly from (9-10) that it yields the same symmetric  conserved quantities
(17-18), though the corresponding $R$-matrix is given in a general form (2)
and with a more involved Lax operator.

Finally, it is important to  observe
 that the Hamiltonian and momentum (9-10) of
the generating model is transformed exactly to the Suris form
(15) under canonical transformation
$(p,q)\rightarrow (P,Q)$ as
\begin {equation}
p =P, \qquad \ q +(\eta -\epsilon(1+c))p =Q
.\end {equation}
%21
Therefore, since   the other models are obtained from the generating
model (9) at different
possible values of the parameters, all of them are  naturally
canonically equivalent  to  Suris model (15). On the other hand
canonical equivalence of the Suris model with the relativistic Toda chain
of Ruijsenaars is already demonstrated in [4].
This  establishes that the whole family of
 different quantum integrable disctrete-time
generalisations of TC presented here
  are canonically equivalent also to the relativistic generalisations
of the Toda chain, though they represent  as such distinct systems
with different Lax operators and $R$-matrices.
\vskip 1 cm
3.~ We switch over now to a recently proposed  integrable  classical
lattice model
close to  TC,  given  in canonical variables $p_i,q_i$
 by the Lax operator
[7]
\begin {equation}
L_{i}(\lambda) = \left( \begin{array}{c}
  \lambda +\gamma p_iq_i+\omega_i
\qquad  \ \ \beta_ip_i
 \\ \alpha_i q_i
\qquad\qquad \ \frac {\alpha_i \beta_i}{\gamma}
          \end{array}   \right),
\end {equation}
%22
and find below a quantum as well as a discrete-time generalisation of it,
 obtained again from the ancestor $L$ operator (3). Consider a
realisation
of (4) with $\theta=0$ as
\begin {eqnarray}
\tau_{1i}^{\pm }=\mp \frac {\hbar\gamma}{2\sin \hbar \eta}
 e^{\mp \eta(N_i+\omega_i)} ,
\ \ \
\tau_{2i}^{\pm }= \frac {\alpha_i \beta_i}{2\gamma}
            e^{\pm \eta N_i}, \ \ \
\tau_{21i}= \beta_iA^+_i,\ \  \  \tau_{12i} = \alpha_i (\cos \hbar \eta ) \
A^-_i,
\end {eqnarray}
%23
where $A^{\pm}$ are generalisation $q$-oscillators [11] given by the
commutation
relations
\begin {equation}
[N_k,A^{\pm}_l]=\pm i \delta_{kl} \hbar \  A^{\pm}_k , \ \
\ A^{-}_k
 A^{+}_l -  e^{\mp 2i \hbar \eta }  A^{+}_l A^{-}_k= i \delta_{kl} \hbar
 \ e^{\pm\eta (2 N+ {\omega- i\hbar})}
.\end {equation}
%24
Through canonical variables these operators may be expressed as
\begin {equation}
N=p q,
\  A^{-}= q f(N), \ \
 A^{+}= f(N) p,
\end {equation}
%25
with
$ f^2(N)= \frac {i \hbar} {2\cos \hbar \eta \ [i\hbar]_\eta}
\frac {1} {N}\left( [2N+\omega- i\hbar]_\eta
-[\omega- i\hbar]_\eta \right)
$ and the notation  $[ x ]_\eta=\frac {\sinh \eta x} {\sinh \eta}$. Inserting
(23) in (3)
and assuming $\xi=e^{\eta \lambda}$
 we get the Lax operator of the discrete-time quantum
model as
\begin {equation}
L_{i}(\lambda) = \left( \begin{array}{c}

\frac { i\gamma \hbar} { [i\hbar]_\eta} [\lambda + N_i +\omega_i]_\eta
\qquad  \ \ \beta_i  A^+_i
 \\ \alpha_i \cos \hbar \eta \ A^-_i
\qquad \ \frac {\alpha_i \beta_i}{\gamma}
       \cosh (\eta(N-\lambda))   \end{array}   \right),
\end {equation}
%26
associated with the quantum $R_0$ matrix (19). In the time-continuous
limit: $\eta \rightarrow 0$, (26) reduces clearly to the form (22),
though giving a quantum version of it with  quantum $R$-matrix:
$R= I+\frac {i \hbar\cal P
} {\lambda-\mu}$, obtainable from (19) at  $\eta \rightarrow 0$.
\vskip 1 cm
4. ~ Though the explicit $R$-matrix and $L$ operators are found for all
the above quantum DTTC, the standard  quantum inverse
scattering method is not applicable to them. The reason of this, as
also  is true for the   TC, is  the absence of pseodovaccum for such
models. However with the use of functional Bethe ansatz (FBA),
Sklyanin was able to transform
the eigenvalue equation of quantum TC from $n$-particle  to a single particle
eigenfunction [3]. It is interesting to observe that the same scheme for
 separation of variables is generalised trivially for the present discrete-time
models. Therefore referring to [3] for details we mention only
its main features sticking to the
$R_0$-matrix  (19). Using QYBE (1) with $L \rightarrow T$ for
 the monodromy matrices $T
(\xi)$ (see (8) for definition) we find
 $[C(\xi),C(\zeta)]=0$, which is crucial for the application of  FBA,
along with functional relations between $C(\xi),A(\zeta)$ and
$C(\xi),D(\zeta)$.
Defining commuting operators $\theta_n$ as $C(\lambda=\theta_n)=0$
and $\xi=e^{i\lambda}, \ \ \xi_n=e^{i\theta_n}$, we construct
conjugate operators $\Lambda^-_j=A(\lambda\rightarrow \theta_j), \ \
\Lambda^+_j=D(\lambda\rightarrow \theta_j)
$
with the proper operator ordering prescription [3]. Parallel to [3]
 the relations
between $C,A$ and $C,D$ show that the operators $\Lambda^{\pm}_j$
shift the operators  $\theta_j$ by $\mp \alpha$. Since
expansion of $C(\xi,\xi_n)$ describes symmetric polynomials in $\xi_n$
and $\Lambda^{\pm}_j$ have shifting effect only on  $\xi_j$,
we achieve a seperation of variable
for the $N$-paricle eigenfunction
$\phi (\xi_1,\xi_2, \ldots ,\xi_N)= \prod_n^N \phi_n(\xi_n)$ symmetric in
$\xi_n$ as
 \begin {equation}
t(  \xi_n) \phi_n(\xi_n)=(A(\xi_n)+D(\xi_n)) \phi_n(\xi_n)=
i^{-N} \phi_n(e^{-i\alpha}\xi_n)+
i^{N} \phi_n(e^{i\alpha}\xi_n)
.\end {equation}
%27
The above described method of seperation of variables is applicable
 directly  to the
discrete-time  models
 associated with $R_0$ matrix (19), while for other models
related to (2) some more effort is needed.

  We have  considered here
the periodic boundary condition  for the discrete-time
family of quantum TC, which corresponds to the root system of $A_{N-1}$.
The extension of these models for other types of boundary conditions
related to other classical algebras would be an interesting problem.

\vskip 1 cm

The author likes to thank Orlando Ragnisco of Rome University for many valuable
discussions and Prof. Vladimir Rittenberg for encouragement and for creating
a stimulating atmosphere in his research group. The support of Alexander
 Humboldt Foundation is thankfully acknowledged.

\newpage
\noindent {\bf  References}
\begin{enumerate}

%1
\item  S.V.~Manakov,   { JETP }  { 67} (1974) 269

 H.~Flaschka and D.W.~McLaughlin,
 { Progr.Theor.Phys. } { 55}  (1976) 438
%2
\item  M.A. Olshanetsky and A.M. Perelomov,
 { Lett.Math.Phys.} { 2 }  (1977) 7

   M.C. Gutzwiller,   { Ann. Phys. } {133} (1981) 304

 V. Pasquier and M. Gaudin,  { J. Phys.} { A 25} (1992) 5243
%3
\item E.K. Sklyanin,   {Lect.Notes.Phys. (Spinger)} 226 (1985) 196

%4
\item Yu.B.~Suris, {  Phys. Lett.} { A 145} (1990) 113
%5

\item S.N.M.~Ruijsenaars {Comm. Math. Phys.}  { 133} (1990) 217
%6
\item A. Kundu A and B. Basu Mallick
 ,{ Mod. Phys. Lett.  }
 { A 7 } ( 1992) 61
%7
\item P.L. Cristiansen, M.F. Jorgensen and V.B. Kuznetsov
,{Lett.Math.Phys.} {29  } (1993) 165
%8
\item  E.K. Sklyanin , L.A. Takhtajan and L.D. Faddeev, Teor. Mat. Fiz.
40
(1979) 194
%9
\item K. Sogo, M. Uchinami, Y. Akutsu and M. Wadati,
 { Prog. Theor. Phys.} { 68 } (1982)  508
%10
\item B.  Basu Mallick  and A. Kundu  { J. Phys.  }  { A 25 } (1992)  4147
%11
\item
A.J. Macfarlane, {  J.Phys. }  {  22 } ( 1989) 4581

 L.C. Biederharn,  { J.Phys.}  { A 22 } ( 1989) L873
\end{enumerate}
\end {document}